\documentclass{aa}
\usepackage{graphicx}
\usepackage{natbib} 

\begin{document}

\def\hii{H~II}
\def\hh{H$_2$}
\def\co{C$^{17}$O}
\def\cs{C$^{34}$S}
\def\hho{H$_2$O}
\def\hhhp{H$_3^+$}
\def\hhhop{H$_3$O$^+$}
\def\hhhsp{H$_3$S$^+$}
\def\hcop{HCO$^+$}
\def\hcsp{HCS$^+$}
\def\hhco{H$_2$CO}
\def\hhcs{H$_2$CS}
\def\hhs{H$_2$S}
\def\chhhoh{CH$_3$OH}
\def\chhhcn{CH$_3$CN}
\def\sotwo{SO$_2$}
\def\ammo{NH$_3$}
\def\amhs{NH$_4$SH}
\def\soo{SO$_2$}
\def\coo{CO$_2$}
\def\oo{O$_2$}
\def\hcch{C$_2$H$_2$}

\def\gtsim{{_>\atop{^\sim}}}
\def\ltsim{{_<\atop{^\sim}}}
\def\vlsr{$V_{\rm LSR}$}
\def\kms{km~s$^{-1}$}
\def\rcm{cm$^{-1}$}
\def\scm{cm$^{-2}$}
\def\ccm{cm$^{-3}$}
\def\mic{$\mu$m}
\def\klm{k$\lambda$}
\def\mjyb{mJy beam$^{-1}$}
\def\msol{M$_{\odot}$}
\def\lsol{L$_{\odot}$}

\def\tkin{$T_{\rm kin}$}
\def\tmb{$T_{\rm MB}$}
\def\trot{$T_{\rm rot}$}
\def\txc{$T_{\rm ex}$}

\def\jms{J.\ Mol.\ Sp.}

\title{Sulphur chemistry in the envelopes of massive young stars}

\titlerunning{Sulphur chemistry around massive young stars}

\author{F.F.S. van der Tak \inst{1} \and A.M.S. Boonman \inst{2}\and
  R. Braakman \inst{2}\and E.F. van Dishoeck\inst{2}} 

\institute{Max-Planck-Institut f\"ur Radioastronomie, Auf dem H\"ugel
  69, 53121 Bonn, Germany \and Sterrewacht, Postbus 9513, 2300 RA
  Leiden, The Netherlands} 

\authorrunning{van der Tak et al.}
\offprints{vdtak@mpifr-bonn.mpg.de}

\date{Received 6 December 2002 / Accepted 4 September 2003}

\abstract{ The sulphur chemistry in nine regions in the earliest
  stages of high-mass star formation is studied through single-dish
  submillimeter spectroscopy.  The line profiles indicate that
  10--50\% of the SO and \soo\ emission arises in high-velocity gas,
  either infalling or outflowing. For the low-velocity gas, excitation
  temperatures are 25~K for \hhs, 50~K for SO, \hhcs, NS and HCS$^+$,
  and 100~K for OCS and \soo, indicating that most observed
    emission traces the outer parts ($T<100$~K) of the molecular
    envelopes, except high-excitation OCS and \soo\ lines.
  Abundances in the outer envelopes, calculated with a Monte Carlo
  program, using the physical structures of the sources derived from
  previous submillimeter continuum and CS line data, are
  $\sim$10$^{-8}$ for OCS, $\sim$10$^{-9}$ for \hhs, \hhcs, SO and
  \soo, and $\sim$10$^{-10}$ for HCS$^+$ and NS. In the inner
  envelopes ($T>100$~K) of six sources, the \soo\ abundance is
  enhanced by a factor of $\sim$100--1000.  This region of hot,
  abundant \soo\ has been seen before in infrared absorption, and must
  be small, $\ltsim$0\farcs2 (180~AU radius).  The derived abundance
  profiles are consistent with models of envelope chemistry which
  invoke ice evaporation at $T\sim 100$~K. A major sulphur carrier in
  the ices is probably OCS, not \hhs\ as most models assume. Shock
  chemistry is unlikely to contribute.
The source-to-source abundance
  variations of most molecules by factors of $\sim$10 do not correlate
  with previous systematic tracers of envelope heating.  Without
  observations of \hhs\ and SO lines probing warm ($\gtsim 100$~K)
  gas, sulphur-bearing molecules cannot be used as evolutionary
  tracers during star formation.

}

\maketitle

\section{Introduction}
\label{sec:intro}

Spectral line surveys at submillimeter wavelengths have revealed
considerable chemical differences between star-forming regions (e.g.,
\citealt{blake87,sutton95,schilke97,helm97,jh98surv,nummel00}; see
\citealt{evd01} for a complete list). While these differences indicate
activity, the dependence of molecular abundances on evolutionary state
and physical parameters is poorly understood. Better insight into this
relation would be valuable for probing the earliest, deeply embedded
phases of star formation where diagnostics at optical and
near-infrared wavelengths are unavailable. This is especially true for
the formation of high-mass stars, for which the order in which
phenomena occur is much less well understood than in the low-mass
case, and for which the embedded phase is a significant fraction
($\approx$10\%) of the total lifetime of $\sim$10$^6$~yr.

Sulphur-bearing molecules are attractive as candidate tracers of early
protostellar evolution \citep{jh98sulf,buck03}.  In models of ``hot
cores'' \citep{char97}, where the chemistry is driven by the
evaporation of icy grain mantles due to protostellar heating, the
abundances of sulphur-bearing molecules exhibit a characteristic
variation with time. After its release from the grain mantles, \hhs\ 
is converted into SO and \soo\ in $\sim$10$^3$~yr, and further into
CS, \hhcs\ and OCS after $\sim$10$^5$~yr. This type of model has had
some success in reproducing the chemistry of star-forming regions
\citep{lang00}.

Besides thermal heating, shocks are potentially relevant to the
chemistry in star-forming regions.  Young stars drive outflows which
expose their surroundings to the effects of shocks. The submillimeter
lines of \soo\ in Orion are very broad, and likely arise in the
outflow \citep{schilke97}.  Models of chemistry behind interstellar
shocks \citep{mitch84,leen88,gpdf93} predict enhancements of \hhs, SO
and \soo\ on time scales of $\sim$10$^4$~yr.  Thus, both by ice
evaporation and by shock interaction, sulphur could act as a clock on
the time scale of $10^4$~yr relevant for the embedded phase of star
formation.

\begin{figure*}[t]
  \begin{center}
\resizebox{\hsize}{!}{\includegraphics[angle=-90]{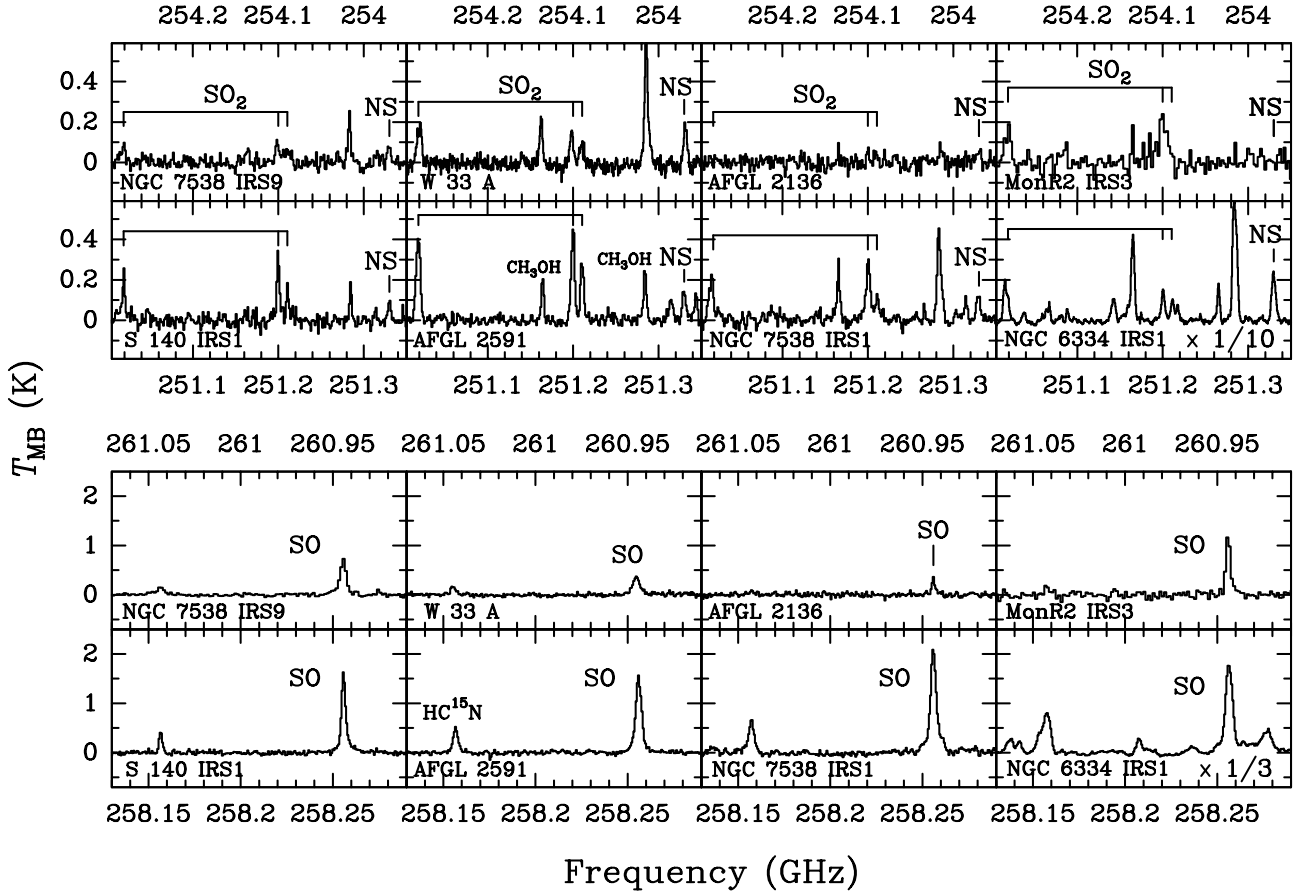}}      
    \caption{Two examples of spectral settings toward eight of our
      sources, showing differences in the relative and in the absolute
      strengths of lines of sulphur-bearing molecules. The top and
      bottom frequency scales are the two receiver sidebands. The data
      for NGC 6334 IRS1 have been divided by 10 and 3 respectively,
      for clarity. The non-labeled lines are due to
      non-sulphur-bearing molecules, of which the strongest are
      labeled in the AFGL~2591 panels.}
    \label{fig:data}
  \end{center}
\end{figure*}

This paper uses submillimeter observations of sulphur-bearing
molecules to test models of chemistry during star formation, and
attempts to order these regions chronologically.  The sources are nine
regions of high-mass star formation with luminosities $1\times 10^4 -
2\times 10^5$~\lsol\ at distances of $1-4$~kpc (Table~\ref{tab:samp}).
Originally selected for their mid-infrared brightness, the sources
were found to have large envelope masses (40-1100~\msol\ within
$r=0.15-0.36$~pc) based on submillimeter data \citep{fvdt00}. Together
with their weak radio emission and their strong molecular outflows,
these masses indicate very early evolutionary stages, probably
preceding the formation of hot cores. Two of our sources may be more
evolved than the others: NGC 6334 IRS1, which has a very rich, `hot
core'-type submillimeter spectrum (\citealt{cutch00}; Thorwirth et
al., in prep.), and NGC 7538 IRS1, which has strong free-free
emission.

Observations of \chhhoh\ lines towards some of these sources
\citep{meth00} indicate an increase in the \chhhoh\ abundance by a
factor of 100 in the warm inner envelope, which is likely due to
evaporation of \chhhoh-rich ices. The excitation of \chhhoh, as well
as that of CO and \hcch, correlates well with other temperature
tracers, such as the gas/solid ratios of \coo\ and \hho, the
45/100~\mic\ colour and the fraction of heated solid $^{13}$\coo\ 
\citep{boog00,fvdt00,mieke:co2}.  Since these quantities also
correlate with the ratio of envelope mass to stellar mass, their
variation likely reflects evolution through the dispersal of the
protostellar envelopes. These findings make the sample a good testbed
for theories of chemical evolution during the earliest deeply embedded
phases of high-mass star formation.

\section{Observations}
\label{sec:obs}

The data presented in this paper are part of a targeted
spectral line survey of embedded massive stars, aimed to track
chemical evolution during the earliest stages of high-mass star
formation. Fifteen frequency settings have been observed, which also
contain many lines of sulphurless molecules. Some of these data have
been published by \citet{fvdt99,meth00,fvdt00}.  These observations
were performed with the 15-m James Clerk Maxwell Telescope
(JCMT)\footnote{The James Clerk Maxwell Telescope is operated by the
Joint Astronomy Centre, on behalf of the Particle Physics and
Astronomy Research Council of the United Kingdom, the Netherlands
Organization for Scientific Research and the National Research Council
of Canada.} on Mauna Kea, Hawaii between 1995 and 1998. The beam size
(FWHM) and main beam efficiency of the antenna were $18''$ and
$64-69$\% at $230$~GHz and $14''$ and $58-63$\% at $345$~GHz. 
The frontends were the receivers A2, B3 and B3i; the backend was the
Digital Autocorrelation Spectrometer, covering $500$~MHz instantaneous
bandwidth. Pointing was checked every 2 hours during the observing and
was always found to be within $5''$. To subtract the atmospheric and
instrumental background, the chopping secondary was used with $180''$
offsets. Total integration times are 30--40 minutes for each frequency
setting. Figure~\ref{fig:data} shows examples of the data towards all
but one of our sources; the data on W3~IRS5 were presented by
\citet{helm97}.

Additional observations of \hhs\ and OCS in the 130-180~GHz window
were carried out with the 30-m telescope of the Institut de Radio
Astronomie Millim\'etrique (IRAM)\footnote{IRAM is an international
  institute for research in millimeter astronomy, cofunded by the
  Centre National de la Recherche Scientifique (France), the Max
  Planck Gesellschaft (Germany), and the Instituto Geografico Nacional
  (Spain).} on Pico Veleta, Spain, in August 2002. The frontend was
the facility receiver C150 and the backend the Versatile Spectral
Assembly (VESPA) autocorrelator. Integration times were 10-20 minutes
using frequency switching with a throw of 7.3~MHz. Data were
calibrated onto \tmb\ scale by multiplying by 1.43, the ratio of
forward and main beam efficiencies. The beam size is $15''$ at these
frequencies.

The \hhs\ 393~GHz line was observed in May 1995 at the 10.4-m Caltech
Submillimeter Observatory (CSO)\footnote{The CSO is supported by NSF
  grant AST 99-80846.}, with the facility receiver as frontend and the
50~MHz AOS as backend. The telescope has a beam size of $25''$ and a
main beam efficiency of 60\% at this frequency.

\begin{table*}[t]
  \begin{center}
    \caption{Source sample.$^a$}
    \label{tab:samp}
    \begin{tabular}[lccccc]{lccccc}
\hline \hline
Source        & R.A.\ (1950)& Dec.\ (1950) & $L$ & $d$ & $N$(\hh) \\
 &(h m s) & (\degr\ \arcmin\ \arcsec) & ($10^{4}$~\lsol) & (kpc) & ($10^{23}$~\scm)\\
\hline   
W3 IRS5       & 02 21 53.1 & +61 52 20 & 17 & 2.2 & 2.3 \\ 
MonR2 IRS3    & 06 05 21.5 &--06 22 26 & 1.3& 0.95 &1.8 \\
NGC 6334 IRS1 & 17 17 32.0 &--35 44 05 & 11 & 1.7 &11.3 \\
W33A          & 18 11 43.7 &--17 53 02 & 10 & 4   & 6.2 \\
AFGL 2136     & 18 19 36.6 &--13 31 40 &  7 & 2   & 1.2 \\
AFGL 2591     & 20 27 35.8 & +40 01 14 &  2 & 1   & 2.3 \\
S140 IRS1     & 22 17 41.1 & +63 03 42 &  2 & 0.9 & 1.4 \\
NGC 7538 IRS1 & 23 11 36.7 & +61 11 51 & 13 & 2.8 & 6.5 \\
NGC 7538 IRS9 & 23 11 52.8 & +61 10 59 &  4 & 2.8 & 3.3 \\
\hline
    \end{tabular}
  \end{center}

$^a$ Luminosity and distance from \citet{henn92} and \citet{gian97}
(MonR2 IRS3) and from \citet{fvdt00} (other sources); \hh\ column
density from this work (MonR2 IRS3) and from \citet{fvdt00} (other
sources).

\end{table*}

\begin{table}
    \caption{Observed transitions}
    \label{tab:lines}
\begin{center}
    \begin{tabular}{lllrcc}
\hline \hline
\noalign{\smallskip}
Molecule   & Transition & Frequency & $E_{\rm up}$ & Telescope & Beam FWHM \\
           &            & (MHz) & (K) & & (arcsec) \\
\noalign{\smallskip}
\hline   

$\rm H_2CS$ &$J_{K_pK_o}=7_{16}\to6_{15}$       &244047.8 & 60.0 & JCMT & 18 \\
$\rm H_2CS$ &$J_{K_pK_o}=10_{19}\to9_{18}$      &348531.9 & 105.2 & JCMT & 14 \\
                                                         
HCS$^+$    & $J=6\to 5$                         &256027.8 & 43.0 & JCMT & 18 \\
HCS$^+$    & $J=8\to 7$                         &341350.8 & 73.7  & JCMT & 14 \\
                                                         
$\rm H_2S$ & $J_{K_pK_o}=1_{10}\to1_{01}$       &168762.8 & 27.9 & IRAM & 15 \\
$\rm H_2S$ & $J_{K_pK_o}=2_{11}\to2_{02}$       &393450.5 & 73.6 & CSO & 25 \\
$\rm H_2S$ & $J_{K_pK_o}=2_{20}\to2_{11}$       &216710.4 & 84.0  & JCMT & 18 \\

NS & $^2\Pi_{1/2}$ $J=11/2\to9/2$               &253969\ $^a$ & 39.9 & JCMT & 18 \\
NS & $^2\Pi_{1/2}$ $J=15/2\to13/2$              &346220\ $^b$ & 71.0 & JCMT & 14 \\

OCS & $J=13\to 12$ & 158107.4 & 53.1 & IRAM & 15 \\
OCS & $J=28\to 27$ & 340449.2 & 237.0 & JCMT & 14 \\
OCS & $J=30\to 29$ & 364749.0 & 271.4 & JCMT & 14 \\

$\rm OC^{34}S$ & $J=18\to 17$ &213553.1 & 97.4 & JCMT & 18 \\
$\rm OC^{34}S$ & $J=22\to 21$ &260991.8 & 144.1 & JCMT & 18 \\

SO & $N_J=5_6 \to 4_5$ & 219949.4 & 35.0 & JCMT & 18 \\
SO & $N_J=6_6 \to 5_5$ & 258255.8 & 56.5 & JCMT & 18 \\
SO & $N_J=6_7 \to 5_6$ & 261843.7 & 47.6 & JCMT & 18 \\
SO & $N_J=8_7 \to 7_6$ & 340714.2 & 81.2 & JCMT & 14 \\
SO & $N_J=8_8 \to 7_7$ & 344310.6 & 87.5 & JCMT & 14 \\
SO & $N_J=8_9 \to 7_8$ & 346528.5 & 78.8 & JCMT & 14 \\

$\rm ^{34}SO$ & $N_J=8_8 \to 7_7$ &337582.2 & 86.1 & JCMT & 14 \\

$\rm SO_2$& $J_{K_pK_o}= 5_{24}\to 4_{13}$& 241615.8 & 23.6 & JCMT & 18 \\
$\rm SO_2$& $J_{K_pK_o}= 5_{33}\to 4_{22}$& 351257.2 & 35.9 & JCMT & 14 \\
$\rm SO_2$& $J_{K_pK_o}= 5_{33}\to 5_{24}$& 256246.9 & 35.9 & JCMT & 18 \\
$\rm SO_2$& $J_{K_pK_o}= 5_{51}\to 6_{42}$& 345149.0 & 75.1 & JCMT & 14 \\
$\rm SO_2$& $J_{K_pK_o}= 6_{33}\to 6_{24}$& 254280.5 & 41.4 & JCMT & 18 \\
$\rm SO_2$& $J_{K_pK_o}= 7_{35}\to 7_{26}$& 257100.0 & 47.8 & JCMT & 18 \\
$\rm SO_2$& $J_{K_pK_o}= 8_{35}\to 8_{26}$& 251210.6 & 55.2  & JCMT & 18 \\
$\rm SO_2$& $J_{K_pK_o}= 9_{37}\to 9_{28}$& 258942.2 & 63.5 & JCMT & 18 \\

$\rm SO_2$& $J_{K_pK_o}=10_{64}\to 11_{57}$     & 350862.7 & 138.8 & JCMT & 14 \\
$\rm SO_2$& $J_{K_pK_o}=11_{1,11}\to 10_{0,10}$ & 221965.2 &  60.4 & JCMT & 18 \\
$\rm SO_2$& $J_{K_pK_o}=11_{39}\to 11_{2,10}$   & 262256.9 &  82.8 & JCMT & 18 \\
$\rm SO_2$& $J_{K_pK_o}=13_{1,13}\to 12_{0,12}$ & 251199.7 &  82.2 & JCMT & 18 \\
$\rm SO_2$& $J_{K_pK_o}=14_{0,14}\to 13_{1,13}$ & 244254.2 &  93.9 & JCMT & 18 \\
$\rm SO_2$& $J_{K_pK_o}=15_{6,10}\to 16_{5,11}$ & 253956.6 & 198.6 & JCMT & 18 \\
$\rm SO_2$& $J_{K_pK_o}=16_{4,12}\to 16_{3,13}$ & 346523.9 & 164.5 & JCMT & 14 \\
$\rm SO_2$& $J_{K_pK_o}=18_{1,17}\to 18_{0,18}$ & 240942.8 & 163.1  & JCMT & 18\\
$\rm SO_2$& $J_{K_pK_o}=18_{4,14}\to 18_{3,15}$ & 338306.0 & 196.8 & JCMT & 14 \\
$\rm SO_2$& $J_{K_pK_o}=19_{1,19}\to 18_{0,18}$ & 346652.2 & 168.1 & JCMT & 14 \\
$\rm SO_2$& $J_{K_pK_o}=20_{1,19}\to 19_{2,18}$ & 338611.8 & 198.9 & JCMT & 14 \\
$\rm SO_2$& $J_{K_pK_o}=22_{2,20}\to 22_{1,21}$ & 216643.3 & 248.4 & JCMT & 18 \\
$\rm SO_2$& $J_{K_pK_o}=24_{2,22}\to 23_{3,21}$ & 348387.8 & 292.7 & JCMT & 14 \\
$\rm SO_2$& $J_{K_pK_o}=25_{3,23}\to 25_{2,24}$ & 359151.2 & 320.9 & JCMT & 14 \\
$\rm SO_2$& $J_{K_pK_o}=28_{2,26}\to 28_{1,27}$ & 340316.4 & 391.8 & JCMT & 14 \\

$\rm ^{34}SO_2$& $J_{K_pK_o}=7_{44}\to 7_{35}$       & 345519.7 & 63.6 & JCMT & 14 \\
$\rm ^{34}SO_2$& $J_{K_pK_o}=10_{46}\to 10_{37}$     & 344245.4 & 88.4 & JCMT & 14 \\
$\rm ^{34}SO_2$& $J_{K_pK_o}=11_{39}\to 11_{2,10}$   & 253936.3 & 81.9 & JCMT & 18 \\
$\rm ^{34}SO_2$& $J_{K_pK_o}=15_{2,14}\to 14_{1,13}$ & 358988.0 & 118.7 & JCMT & 14 \\
$\rm ^{34}SO_2$& $J_{K_pK_o}=19_{1,19}\to 18_{0,18}$ & 344581.1 & 167.4 & JCMT & 14 \\
$\rm ^{34}SO_2$& $J_{K_pK_o}=29_{9,21}\to 30_{8,22}$ & 257466.7 & 590.7 & JCMT & 18 \\

\noalign{\smallskip} 
\hline                
\end{tabular}
\end{center}

$^a$ Blend of the $F=7\to 6$, $F=6\to 5$ and $F=5\to 4$ hyperfine components 

$^b$ Blend of the $F=9\to 8$, $F=8\to 7$ and $F=7\to 6$ hyperfine components 

\end{table}

Reduction was carried out using the CLASS package developed at IRAM.
Linear baselines were subtracted and the spectra were smoothed once
and calibrated onto \tmb\ scale. Only the frequency switched 30m data
required higher order polynomials to fit the baseline. The final
spectra have a resolution of $0.3-1.5$~\kms\ and rms noise levels (in
\tmb) of $20-30$~mK for the 150~and 230~GHz bands, and $30-50$~mK for
the 345~GHz band. Multiplying these values by the line width
(Table~\ref{tab:velo}) yields the uncertainties of the line fluxes in
Table~\ref{tab:flux}. Although the absolute calibration is only
correct to $\approx$30\%, the relative strength of lines within one
frequency setting is much more accurate.

\begin{table*}[ptb]
\begin{center}
\caption{Observed line fluxes $\int T_{\rm{MB}} dV$ (K km s$^{-1}$)
  from Gaussian fits, and 1$\sigma$ upper limits. Dots denote missing data.}
\label{tab:flux}
\begin{tabular}{lllrrrrrrrr}
\hline \hline
\noalign{\smallskip}
Transition & W3$^a$ &AFGL &\multicolumn{2}{c}{NGC 7538} &MonR2 &S140 &NGC6334 &AFGL  & W33A\\
           & IRS5   &  2591  &IRS1      &IRS9           &IRS3  &IRS1 &IRS1    & 2136 &     \\
\noalign{\smallskip}
\hline
\noalign{\smallskip}

$\rm H_2CS$ $7_{16}\to6_{15}$  &     0.8 &1.4 &3.8 &1.6 &0.9 &1.5 &10.6 &0.3 &2.9\\
$\rm H_2CS$ $10_{19}\to9_{18}$ &$<$0.3   &1.1 &3.5 &1.2 &0.3 &0.8 &...  &1.0 &3.6\\

HCS$^+$  $6\to5$             &0.3      &... &... &... &...&... &...  &... &...\\
HCS$^+$  $8\to7$             &0.3      &0.6 &1.6 &0.8 &...    &0.7 &16.6 &0.3 &1.1\\

$\rm H_2S$ $1_{10}\to1_{01}$ & 9.4 &13.1&15.4&7.7 & 7.6   &14.3&...  &6.7 &17.8\\
$\rm H_2S$ $2_{11}\to2_{02}$ & 2.6 &... &1.6 &... &   ... &... &...  &   ... &...\\
$\rm H_2S$ $2_{20}\to2_{11}$ & 2.2 &0.8 &1.9 &0.4 &$<$0.2 &0.6 &22.8 &$<$0.1 &1.9\\

NS $11/2\to9/2$            &... &0.9    &0.9    &0.5 &$<$0.4 &0.5 &15.9 &$<$0.2 &2.2\\
NS $15/2\to13/2$           &... &$<$0.8 &$<$1.2 &... &... &... &11.6 &... & 2.8\\

OCS $13\to 12$ &      ... & 0.4 &0.8  &0.3   &... &$<$0.3&...&...&...\\
OCS $28\to 27$ &      0.9 & 0.3 &...  &...   &... &... &...&...&2.6\\
OCS $30\to 29$ & $<$0.4   & 0.3 &...  &...   &... &... &...&...&...\\

$\rm OC^{34}S$ $18\to 17$ &     ... &...  &...   &... &... &... & 4.8 &...&0.4\\
$\rm OC^{34}S$ $22\to 21$ &$<$0.1   &...  &...   &... &... &... &11.6 &...&...\\

SO $5_6 \to 4_5$ &     17.4 &9.9 &8.4 &1.3 &3.7 &10.8 &... &2.0&3.0\\
 broad           &     11.2 &4.5 &6.5 &5.2 &3.6 &3.7  &... &1.2&5.1\\
SO $6_6 \to 5_5$ &     26.1 &5.3 &5.8 &2.5 &3.5 &4.0  &26.1$^c$ &0.4&3.1$^c$\\
 broad           &     22.1 &2.5 &5.7 &1.5 &1.4 &2.6  &...      &0.6&...\\
SO $6_7 \to 5_6$ &     14.7 &7.3 &10.0 &3.0 &4.8 &5.7 &... &1.5&3.5\\
 broad           &     13.2 &3.7 & 7.3 &3.0 &3.1 &4.7 &... &1.3&5.6\\
SO $8_7 \to 7_6$ &     17.9 &3.7 &6.5$^b$ &1.3 &... &2.6 &14.5 &$<$0.2&1.8$^c$\\
  broad          &     26.8 &5.1 &...     &2.4 &... &5.5 &23.4 &$<$0.2&...\\
SO $8_8 \to 7_7$ &     31.7 &3.4 &9.5 &4.7$^b$ &... &2.6 &47.3$^c$ &1.5&4.1\\
  broad          &     28.8 &4.5 &6.7 &..      &... &5.0 &...      &0.5&6.4\\
SO $8_9 \to 7_8$ &     27.9 &5.6 &8.3 &1.5 &3.6 &3.9 &26.4 &1.4$^c$&2.2\\
  broad          &     58.1 &5.1 &5.9 &3.9 &5.7 &6.0 &34.4 &...    &6.7\\

$\rm ^{34}SO$   $8_8 \to 7_7$ &    18.9 &2.7 &0.9 &$<$0.3 &... &$<$0.4 &19.6 &...&$<$0.7\\

$\rm SO_2$ $5_{24}\to 4_{13}$&      8.2 &2.3 &2.0 &0.7 &1.0 &1.1   &6.8$^d$ &...&...\\
$\rm SO_2$ $5_{33}\to 4_{22}$&     12.2 &... &... &... &... &...   &... &...&2.9\\
$\rm SO_2$ $5_{33}\to 5_{24}$&      6.9 &... &... &... &... &$<$1  &... &...&...\\
$\rm SO_2$ $5_{51}\to 6_{42}$& $<$0.3   &... &... &... &... &...   &7.1 &...&...\\
$\rm SO_2$ $6_{33}\to 6_{24}$&     20.8 &3.0 &1.7 &0.6 &1.6 &1.1   &12.1&$<$0.1&2.2\\
$\rm SO_2$ $7_{35}\to 7_{26}$&     10.2 &1.7 &1.1 &$<$0.4&0.7 &0.5 &... &...&1.0\\
$\rm SO_2$ $8_{35}\to 8_{26}$&     12.7 &1.7 &0.9 &0.4 &0.7 &  0.7 &5.9 &$<$0.2&1.4\\
$\rm SO_2$ $9_{37}\to 9_{28}$&      9.6 &2.4 &1.5 &$<$0.2 &... &$<$0.5 &... &...&3.7\\
$\rm SO_2$ $10_{64}\to 11_{57}$    &      3.4 &1.7 &$<$0.4&... &... &... &... &...&0.8\\
$\rm SO_2$ $11_{1,11}\to 10_{0,10}$&      ... &... &...   &... &... &... &... &0.3&...\\
$\rm SO_2$ $11_{39}\to 11_{2,10}$  &      ... &... &...   &... &... &... &... &...&1.6\\
$\rm SO_2$ $13_{1,13}\to 12_{0,12}$&     19.0 &2.7 &2.2   &0.7&1.5  &1.5 &8.8 &$<$0.2&1.7\\
$\rm SO_2$ $14_{0,14}\to 13_{1,13}$&     12.2 &2.8 &1.7   &... &1.3 &1.3 &... &0.3&1.9\\
$\rm SO_2$ $15_{6,10}\to 16_{5,11}$&      ... &0.6 &...   &... &... &... &... &...&...\\
$\rm SO_2$ $16_{4,12}\to 16_{3,13}$&     29.5 &... &...   &... &... &... &... &0.6&2.0\\
$\rm SO_2$ $18_{1,17}\to 18_{0,18}$&      7.7 &1.0 &...   &... &... &... &... &...&0.7\\
$\rm SO_2$ $18_{4,14}\to 18_{3,15}$&     13.3 &2.9 &...   &... &1.2 &... &... &...&1.3\\
$\rm SO_2$ $19_{1,19}\to 18_{0,18}$&     31.7 &2.6 &$<$0.4&$<$0.4&...&$<$0.7&7.0 &0.4&...\\
$\rm SO_2$ $20_{1,19}\to 19_{2,18}$&     16.0 &2.7 &...   &... &... &... &... &...&...\\
$\rm SO_2$ $22_{2,20}\to 22_{1,21}$&      5.5 &0.8 &...   &... &... &... &1.7 &...&0.5\\
$\rm SO_2$ $24_{2,22}\to 23_{3,21}$&      6.1 &3.2 &$<$0.6&... &... &... &... &...&...\\
$\rm SO_2$ $25_{3,23}\to 25_{2,24}$&      4.1 &2.8 &1.4   &... &... &... &... &...&1.0\\
$\rm SO_2$ $28_{2,26}\to 28_{1,27}$&     11.3 &... & ...  &... &... &... &... &...&0.6\\

$\rm ^{34}SO_2$ $7_{44}\to 7_{35}$      &      2.4 &1.0 &...   &... &... &... &... &...&...\\
$\rm ^{34}SO_2$ $10_{46}\to 10_{37}$    &      2.5 &1.1 &...   &... &... &... &... &...&...\\
$\rm ^{34}SO_2$ $11_{39}\to 11_{2,10}$  &      ... &1.1 &...   &... &... &... &... &...&0.8\\
$\rm ^{34}SO_2$ $15_{2,14}\to 14_{1,13}$&      0.7 &1.1 &...   &... &... &... &... &...&0.8\\
$\rm ^{34}SO_2$ $19_{1,19}\to 18_{0,18}$&      5.8 &1.2 &$<$0.4&$<$0.4&... &$<$0.4 &... &...&...\\
$\rm ^{34}SO_2$ $29_{9,21}\to 30_{8,22}$& $<$0.2   &... &...   &... &... &... &... &...&0.3\\

\noalign{\smallskip}
\hline
\noalign{\smallskip}
\end{tabular}
\end{center}

$^a$ From \citet{helm97}, except SO

$^b$ Attributed to envelope based on center velocity

$^c$ Attributed to envelope based on line width

$^d$ Difficult fit due to partial blend with other lines

\end{table*}

Line identification was performed using the JPL catalog
\citep{pick98}\footnote{http://spec.jpl.nasa.gov}.  A matching
frequency is the main criterion for assignment, but the upper energy
level of the transition and the complexity of the molecule are also
considered. Table~\ref{tab:lines} has details about the observed
transitions. 

For each detected feature, a satisfying match could be found in the
catalog within 0.5~MHz, except for the \hhcs\ $10_{19}-9_{18}$ line at
348.5~GHz, whose residual is 2~MHz. Laboratory spectroscopy of \hhcs\ 
only exists up to 250~GHz \citep{beers72}, and the analysis in the JPL
catalog only includes pure rotation and centrifugal distortion terms.
At higher frequencies and for higher $J$-values, higher order terms
become important, which may be why the catalog prediction is
inaccurate for this line. Our data indicate a frequency of
348534.2~MHz.

\section{Results}
\label{sec:res}

\subsection{Line profiles}
\label{sec:profi}

\begin{figure}[t]
  \begin{center}
\resizebox{\hsize}{!}{\includegraphics[angle=-90]{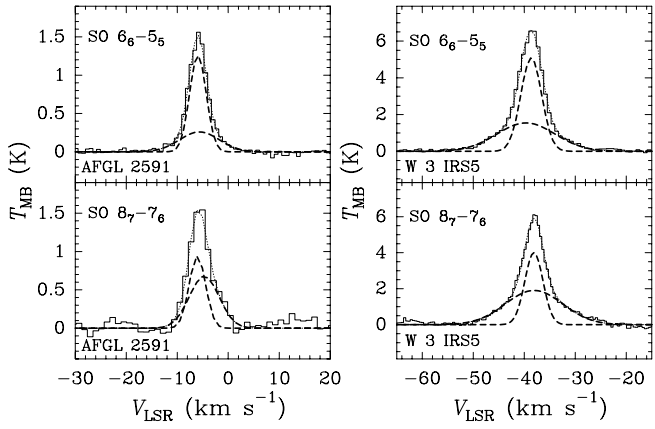}}    
    \caption{Observed velocity profiles of the SO 6$_6$--5$_5$ and
      8$_7$--7$_6$ lines in the sources AFGL~2591 and W3~IRS5,
      showing the low- and high-velocity components. The dashed lines
      are our fits to the individual components and the dotted line is
      the sum of the two components.}
    \label{fig:2comp}
  \end{center}
\end{figure}

We have fitted the detected lines with Gaussian profiles.
Table~\ref{tab:flux} reports the observed fluxes and
Table~\ref{tab:velo} the widths of the lines.  While most lines could
be satisfactorily fitted with a single Gaussian, the lines of SO,
which are the strongest lines that we have observed, required two
Gaussians for a good fit to their profiles (Fig.~\ref{fig:2comp};
Table~\ref{tab:velo}).  One component agrees in position and width
with the values measured for C$^{17}$O and C$^{34}$S by \citet{fvdt00}
and is attributed to a static envelope.  Table~\ref{tab:cs_lines}
reports similar measurements for the source MonR2 IRS3 which van der
Tak et al.\ did not study.  The other component has $2.8$ times as
large a width on average. This high-velocity emission may trace
outflowing or infalling motions.  The contributions from high- and
low-velocity gas to the SO line fluxes are about equal, unlike CS,
where the envelope carries 90\% of the fluxes (\citealt{fvdt00};
Table~\ref{tab:cs_lines}).  The infrared absorption and submillimeter
emission lines of CO towards these sources also show multiple velocity
components \citep{mitch91:episode,mitch91:irs9,mitch92}.  However, the
velocities do not match in detail, as expected because of differences
in spectral and angular resolution between infrared and submillimeter
data on one hand, and excitation differences between CO and SO on the
other.

The fit of two Gaussians to the SO lines is excellent, which is not
always the case for CO. An alternative method is to use area
integrals, which do not attribute any low-velocity gas to the
high-velocity component. Our approach may overestimate the
high-velocity contribution to the line flux by factors up to 2.

The ratios of the high- to low-velocity contributions to the SO lines
are $\ltsim 1$ for the lines in the 230~GHz band, and $\gtsim 1$ for
the lines in the 345~GHz band. This difference could be due to the
higher temperatures and densities required for excitation of the
345~GHz lines, or to the smaller beams in which they are observed.
Since the high-velocity wings on the CS lines in these sources also
become more pronounced in smaller beams, independent of transition
\citep{fvdt00}, the broad component probably is more compact than the
envelope, but not necessarily warmer or denser. The high-velocity gas
in these sources thus seems to be confined to small radii. This
situation differs from that in molecular clouds, where line width
increases with size, and from that in many molecular outflows, where
velocities usually increase with radius (e.g., \citealt{lee02}).
However, while CO data show arcminute-sized outflows in our sources,
higher velocities naturally occur at smaller radii in the case of
infall. For example, gas at 1,000 AU (1$''$ at 1~kpc) in free fall
onto a 10~\msol\ object would have $V_{\rm inf}=4$ \kms, consistent
with the observed width of the broad component (Table~\ref{tab:velo}).

The broad component peaks at slightly more blueshifted velocities (by
$0.17$~\kms\ on average) than the envelope, a situation found before
for CS and \hhco\ \citep{fvdt00}. In the case of outflow, obscuration
by $A_V=10^4$~mag of dust could explain the absence of redshifted
wings. Since some redshifted SO is seen (Fig.~\ref{fig:2comp}), the
requirement relaxes to several 1000~mag. Of the sources discussed
here, only NGC 6334 IRS1 may have such a large amount of dust in its
envelope (Table~\ref{tab:samp}); the other sources may need
circumstellar disks seen face-on.  In the case of infall, a mixture of
absorption and emission is expected at redshifted velocities
\citep{choi02}, which may partly cancel each other if unresolved, to
give the impression of predominantly blueshifted emission.  We
conclude that both infall and outflow are plausible explanations for
the high-velocity gas, and recommend interferometric observations of
SO to decide between these options.

For most sources in our sample, the width of the \soo\ lines is
between the values for the low- and high-velocity components.  The
only source where the \soo\ lines are strong enough to fit two
components is W3~IRS5, where the line profiles do not show wings.
However, most of these data, taken from \citet{helm97}, were observed
with an AOS as backend, and have lower spectral resolution than our
data. It seems therefore plausible that, like for SO, about half of
the \soo\ emission arises in high-velocity gas.

For the other molecules, the signal to noise ratio is not high enough
to disentangle multiple velocity components. This paper assumes that
these lines originate in low-velocity gas, which given our results for
SO may overestimate column densities and abundances
(\S\S~\ref{sec:cold}, \ref{sec:abund}) by factors of up to 2.

\begin{table*}[hbtp]
\begin{center}
\caption{Widths $\Delta V$ (\kms) of the emission line components,
  averaged over transition. Numbers in brackets denote uncertainties
  in units of the last decimal, except when only one transition was
  observed. Dots denote missing data. For SO, values for the narrow and the broad components are
  reported, whose positions are also given.}
\label{tab:velo}
\begin{tabular}{lrrrrrrrrrrr}
\hline \hline
\noalign{\smallskip}
Source &\multicolumn{2}{c}{CS} &\multicolumn{2}{c}{SO (narrow)}  &\multicolumn{2}{c}{SO (broad)} & \soo & OCS & \hhs & \hhcs & HCS$^+$  \\
       &\multicolumn{2}{c}{\hrulefill} &\multicolumn{2}{c}{\hrulefill}    &\multicolumn{2}{c}{\hrulefill} \\
                    &\vlsr & $\Delta V$ &\vlsr & $\Delta V$  &\vlsr      &   $\Delta V$ \\
\noalign{\smallskip}
\hline
\noalign{\smallskip}
W3 IRS5          & $-$38.4(3) &  2.7(11)   & $-$38.7(5) &  4.8(3) & $-$39.0(10) & 13.5(14) & 6.6(12) & 3.5(20)   & 5.6(6) & 2.9    & 1.1(4)\\
W33A             & $+$37.5(9) &  5.4(5)    & $+$37.8(3) &  3.6(2)  & $+$36.9(10) & 11.6(11) & 5.0(3)  &   ...     & 2.5    & 2.2(7) & ...   \\
AFGL 2136        & $+$22.8(1) &  3.1(4)    & $+$22.9(2) &  2.5(8)  & $+$23.6(16) &  6.5(13) & 5.0(7)  & 4.1(2)$^a$& 5.4    & 6.0    &  5.4 \\
AFGL 2591        &  $-$5.5(2) &  3.3(6)    &  $-$6.0(3) &  3.7(2)  &  $-$5.6(9)  &  8.6(15) & 5.0(30) & 4.50      & 4.1    & 5.1(12)&  7.1 \\
S140 IRS1        &  $-$7.0(2) &  3.3(3)    &  $-$7.0(1) &  2.5(2)  &  $-$8.1(2)  &  9.6(22) & 1.9(10) &   ...     & 3.6    & 2.9(8) &  1.0 \\
NGC 7538 IRS1    & $-$57.4(5) &  4.1(14)   & $-$57.5(4) &  3.8(4) & $-$58.4(6)  & 11.5(39) & 5.1(9)  & 3.8(1)    & 3.8(2) & 3.2(4) &  3.2  \\
NGC 7538 IRS9    & $-$57.2(3) &  4.1(5)    & $-$57.4(3) &  3.2(8)  & $-$55.9(5)  &  7.8(21) & 3.5(8)  &   ...     & 2.9(1) & 2.9    &  2.6  \\
MonR2 IRS3       & $+$10.2(1) &  2.4(2)    & $+$10.1(1) &  3.0(1)  & $+$10.0(16) &  8.2(4)  & 7.0(9)  & 3.1       & 4.0(5) & 4.4    &  3.6  \\
NGC 6334 IRS1    &  $-$7.4(2) &  5.3(3)    & $-$7.2(11) &  4.1(1)  &  $-$7.4(13) &  11.1(1) & 5.2(10) & 2.1       & 4.1(3) & 3.5(1) &  3.5  \\
\noalign{\smallskip}
\hline
\end{tabular}
\end{center}

$^a$ Value for OC$^{34}$S

\end{table*}

\begin{table}[bhtp]
  \begin{center}
    \caption{Observations of CS, C$^{34}$S and C$^{17}$O lines towards MonR2 IRS3.}
    \label{tab:cs_lines}
\begin{tabular}{cccccc}
\hline \hline
\noalign{\smallskip}
 Line & Frequency &\tmb&$\Delta V$ &$\int T_{\rm MB} dV$\\
      & (MHz)     &(K) &(km s$^{-1}$) &(K km s$^{-1}$)\\
\noalign{\smallskip}
\hline
\noalign{\smallskip}
CS $5 \to 4$ &244935.6 &9.2 &2.7 &27\\
               &broad  &0.2 &8.0 &1.9\\
CS $7 \to 6$ &342883.0 &6.6 &2.5 &17.8\\
               &broad  &0.3 &8.0 &2.4\\
CS $10\to 9$ &489751.0 &2.0 &2.7 &5.6\\

C$^{34}$S $5 \to 4$ &241016.2 &0.6 &2.3 &1.4\\
C$^{34}$S $7 \to 6$ &337396.7 &0.4 &2.6 &1.1\\

C$^{17}$O $3 \to 2$ &337060.9 &2.3 &2.6 &6.4\\
\noalign{\smallskip}
\hline
    \end{tabular}
  \end{center}
\end{table}

\subsection{Excitation temperatures}
\label{sec:texc}

For SO and \soo, the number of detected lines is large enough to
construct rotation diagrams. This method, described in detail by
\citet{blake87} and \citet{helm94}, assumes that the lines are
optically thin and that the molecular excitation can be described by a
single temperature, the rotation temperature. If radiative decay
competes with collisional excitation, the rotation temperature
lies below the kinetic temperature. If the lines are optically thick,
the rotation diagram underestimates the column density and
overestimates the excitation temperature. Another complication is that
the lines are measured in beams of unequal size, so that somewhat
different volumes of gas are probed. Despite these caveats, the
rotation diagram provides a useful first estimate of excitation
conditions and molecular column densities, and a stepping stone
towards a more sophisticated analysis (\S~\ref{sec:abund}).

\begin{table}[bhtp]
  \begin{center}
    \caption{Rotation temperatures (K). Numbers in brackets denote uncertainties.}
    \label{tab:trot}
    \begin{tabular}{lrrrr}
\hline \hline
\noalign{\smallskip}
Source        & \multicolumn{2}{c}{SO} &\multicolumn{1}{c}{\soo} & \chhhoh$^c$ \\
        & \multicolumn{2}{c}{\hrulefill} & \\
              & Narrow & Broad &          \\
\noalign{\smallskip}
\hline   
\noalign{\smallskip}
W3 IRS5       & $>$100 & $>$100  & 154(14)$^a$ &  73(32)\\ 
W33A          & 54(18) &  96(64) & 124(12)     & 155(19)\\
AFGL 2136     & 59(24) &  34(9)  & $>$200      & 143(12)\\
AFGL 2591     & 31(6)  & $>$100  & 185(28)$^b$ & 163(9) \\
S140 IRS1     & 26(4)  & $>$100  &  56(16)     & 41(4)  \\
NGC 7538 IRS1 & 65(27) &  61(25) &  47(12)     & 189(30)\\
NGC 7538 IRS9 & 48(17) &  45(15) &  ...        & 29(2)  \\
MonR2 IRS3    & 47(20) & $>$100  & 125(35)     & 110(20)\\
NGC 6334 IRS1 & 75(72) &  ...    &  45(15)     & 213(37)\\
\noalign{\smallskip}
\hline
    \end{tabular}
  \end{center}

$^a$ Better fit with two components: \trot = 70(14) \& 254(20) K.

$^b$ Better fit with two components: \trot = 47(11) \& 197(67) K.

$^c$ From JCMT data \citep{meth00,boonman99}.

\end{table}

Table~\ref{tab:trot} presents the results. In the case of SO, separate
fits were made to the low- and high-velocity components.  For NGC 7538
IRS9, the detected \soo\ lines do not cover a large enough range of
energy levels to constrain \trot. The derived temperatures are
30...70~K for SO and 50...190~K for \soo, reflecting the available
range of energy levels.  Lower limits to \trot\ are poor fits to the
data, probably caused by optically thick lines. Indeed, the
SO/$^{34}$SO and \soo/$^{34}$\soo\ line ratios of $1.3$...$\gtsim 10$
are below the isotopic abundance ratio of $^{32}$S/$^{34}$S=22, which
indicates optical depths of a few in the SO and \soo\ lines.

\begin{figure}[htbp]
  \begin{center}
\resizebox{\hsize}{!}{\includegraphics{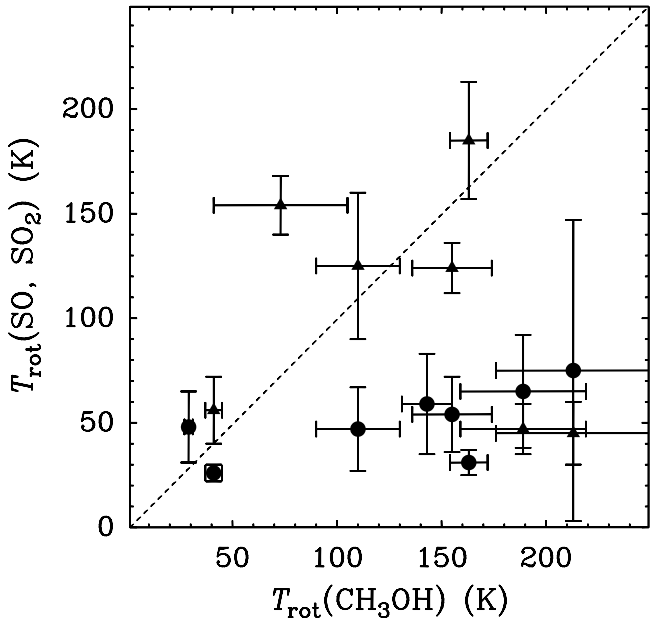}}        
    \caption{Rotation temperatures of SO (dots) and \soo\ (triangles)
      versus values for \chhhoh.}
    \label{fig:trot}
  \end{center}
\end{figure}

The last column of Table~\ref{tab:trot} lists rotation temperatures of
\chhhoh, measured previously. Strictly, these numbers are only lower
limits to the kinetic temperatures in a $15''$ beam, but they show
good correlation with \txc(\hcch) which does directly trace the
kinetic temperature \citep{meth00}. The rotation temperatures of SO
and \soo\ are not clearly correlated with \trot\ (\chhhoh) within the
uncertainties (Fig.~\ref{fig:trot}). Therefore we cannot conclude at
this point whether these molecules trace the same gas within the
envelopes. In any case, differences in beam sizes and optical depth
effects in the lines make it hard to draw firm conclusions before a
more detailed radiative transfer analysis is presented in
\S~\ref{sec:abund}.

For the other molecules, for which only few lines are available, the
line ratios have been used to constrain \txc.  Source-averaged line
ratios indicate \txc$\approx$50~K for \hhcs, NS and HCS$^+$,
\txc$\approx$25~K for \hhs\ and \txc$\approx$100~K for
OCS. These values are only rough estimates, especially if
based on few sources and/or lines.

\subsection{Column densities}
\label{sec:cold}

Table~\ref{tab:cold} presents beam-averaged molecular column
densities, estimated from the observed line fluxes using the
excitation temperatures found above. The column densities for \hhcs\ 
and \hhs\ are the sums of the ortho and para species, assuming an
ortho to para ratio of~3. After calibration, the adopted \txc\ is the
main source of uncertainty in the column densities, especially if only
one or two lines have been observed. For example, decreasing \txc\ 
from 50 to 25~K increases column density estimates by factors of
$\approx$2, while increasing \txc\ from 50 to 75~K decreases column
density estimates by $\ltsim$20\%. In addition, column densities may
have been underestimated by factors of a few due to nonnegligible
optical depth.

\begin{table*}[hbtp]
  \begin{center}
    \caption{Column densities (\scm) in 15--20$''$ beams.}
    \label{tab:cold}
    \begin{tabular}[lrrrrrrrrr]{lrrrrrrrrr}
\hline \hline
\noalign{\smallskip}
Source & \multicolumn{2}{c}{SO ($10^{13}$)} & \soo   & OCS & \hhs & \hhcs  & NS & HCS$^+$  \\
       & \multicolumn{2}{c}{\hrulefill} & & & & & & \\
       &Narrow & Broad       &$10^{13}$&$10^{13}$&$10^{13}$&$10^{12}$&$10^{12}$&$10^{11}$ \\
\noalign{\smallskip}
\hline   
\noalign{\smallskip}
NGC 7538 IRS9 &  3 &    4 &...    &0.6   &  2 &  6 & 1  &  5 \\
S140 IRS1     & 10 & $>$6 &  7    &$<$0.8&  3 &  7 & 1  &  6 \\
W3 IRS5     &$>$27 &$>$11 &530$^a$& 2    &  7 &  5 & ...&  3 \\ 
MonR2 IRS3    &  5 & $>$5 & 15    &...   &  3 &  4 &$<$3& ...\\
AFGL 2136     &  2 &    1 & $>$3  &...   &  2 &  4 &$<$2&  3 \\
W33A          &  4 &    9 & 26    & 9    &  4 & 10 & 4  &  6 \\
AFGL 2591     & 10 & $>$7 & 52$^b$& 1    &  3 &  6 & 2  &  5 \\
NGC 7538 IRS1 & 12 &    9 & 11    & 2    &  5 & 15 & 2  & 11 \\
NGC 6334 IRS1 & 40 &  ... & 46    &24    & 60 & 30 & 20 & 84 \\
\noalign{\smallskip}
\hline
    \end{tabular}
  \end{center}

$^a$ Better fit with two components: $N$=1.0 and 2.0 $\times 10^{15}$~\scm.

$^b$ Better fit with two components: $N$=1.6 and 5.5 $\times 10^{14}$~\scm.

\end{table*}

To search for trends of column densities with temperature,
Table~\ref{tab:cold} lists the sources in order of increasing
$T$(\chhhoh). The only trend seems to be that the
warmest source has the largest column density of all molecules, except
SO and \soo.  In order to investigate whether these trends in the
column density reflect chemical differences between sources or are due
to optical depth or beam size effects, a more detailed radiative
transfer analysis is presented in \S~\ref{sec:abund}.

\subsection{Comparison with infrared data}
\label{sec:ir}

\citet{kean01} observed the $\nu_3$ band of \soo\ around 7.3~\mic\ in
absorption towards five of the sources studied here, and found column
densities of \soo\ of a few $10^{16}$~\scm, a factor of $\sim$100
higher than the submillimeter data indicate.  In contrast, infrared
observations of CO and dust yield 3--5 times {\em lower} column
densities than submillimeter data, as a result of non-spherical
geometry on scales $\ltsim 1''$ \citep{fvdt00}.  Correcting for
optical depth in the \soo\ submillimeter lines increases the column
density estimate, but only by factors of a few (\S~\ref{sec:texc}),
which does not explain the discrepancy with the infrared values.  More
likely to be important are chemical effects on small scales. The
sources are centrally concentrated, $n$(\hh)$\propto r^{-\alpha}$ with
$\alpha=1.0-1.5$, so that absorption data are more sensitive to warm,
dense gas at small radii, while emission data probe more extended,
cooler and less dense gas. The infrared excitation temperatures of
225--750~K also suggest that the \soo\ absorption arises in warm gas.

The infrared estimates of $N$(\soo) imply optically thick
submillimeter lines of \soo, but their brightness measured with the
JCMT is much lower than the infrared \txc.  The implied limit on the
source size is $\theta_S = 15'' (T_{\rm ex} / T_0)^{-0.5}$, with $T_0$
the typical \tmb\ of 0.1~K.  Inferred values of $\theta_S$ range from
0\farcs17 for AFGL~2591 to 0\farcs32 for MonR2 IRS3, corresponding to
linear radii of $100-250$~AU. The limited spectral resolution of the
infrared data prohibits assignment of the absorbers to either velocity
component in emission, which however does not affect our conclusion.

\section{Radiative transfer analysis}
\label{sec:abund}

\subsection{Model description}

To estimate molecular abundances we have used the Monte Carlo
radiative transfer program by \citet{hst00}
\footnote{http://talisker.as.arizona.edu/$\sim$michiel/ratran.html}. 
Starting from radial profiles of
the density and kinetic temperature, this program solves for the
molecular excitation as a function of radius. Besides collisional
excitation, radiation from the cosmic microwave background and thermal
radiation from local dust are taken into account. The result is
integrated over the line of sight and convolved with the appropriate
telescope beam. Observed and synthetic line fluxes are compared with a
$\chi^2$ statistic to find the best-fit abundance. Initially,
abundances were assumed to be constant with radius.

Molecular data were mostly taken from the database by Sch\"oier et
al.\ (in prep.)\footnote{http://www.strw.leidenuniv.nl/$\sim$moldata},
which paper also describes our extension of the \soo\ rate
coefficients by \citet{green95} to higher energy levels. No rate
coefficients are available for NS, so its excitation was assumed to be
thermalized at the temperature of each grid point of the model.  The
population of the $^2\Pi_{3/2}$ state of NS, which lies 225~K above
the $^2\Pi_{1/2}$ state, was assumed to be negligible.  In the case of
\hhs, \citet{ball99} measured the $1_{10}-1_{01}$ rate coefficient in
collisions with He at $T=1.36-35.3$~K. Our model uses the rate
coefficients by \citet{green93} for \hho--He, multiplied by 2.6 based
on comparison with the Ball et al.\ data, and scaled for the different
reduced mass of the \hhs--\hh\ system. \citet{turn96} lists reasons
why \hhs\ rate coefficients are likely to be larger than \hho\ values.


\subsection{Physical structure}

The temperature and density structures of all our sources but MonR2
IRS3 were derived by \citet{fvdt00}. We have followed the same
procedure to infer the structure of MonR2 IRS3, assuming a power-law
density structure $n=n_0(r/r_0)^{-\alpha}$. The temperature profile is
derived from the observed luminosity of $1.3\times 10^4$~\lsol\ 
\citep{henn92}, while $n_0$ is constrained by the submillimeter
photometry of \citet{gian97}. These dust models solve the grain
heating and cooling self-consistenly, using opacity model~5 from
\citet{oss94}. These ``OH5'' opacities are the only ones that yield
dust masses consistent with C$^{17}$O measurements of AFGL~2591, where
CO depletion is known to be negligible \citep{fvdt99}.  To constrain
$\alpha$, the CS and C$^{34}$S line spectrum of MonR2 IRS3 was modeled
with the Monte Carlo program. Besides our own JCMT measurements
(Table~\ref{tab:cs_lines}), data from \citet{taf97} and \citet{choi00}
were used. The best fit was obtained for $\alpha=1.25$
(Table~\ref{tab:monr2}), with a reduced $\chi^2=2.33$ over 11 degrees
of freedom, although $\alpha=1.0$ ($\chi^2=3.1$) and $\alpha=1.5$
($\chi^2=3.8$) also give acceptable fits. Radial profiles of 350~\mic\ 
dust emission for our sources indicate larger values of $\alpha$, by
up to $0.5$, than CS line emission \citep{kaisa02}. If this difference
holds more generally, it may reflect a reduced CS abundance at small
radii, although the models of \S~\ref{sec:chem_mdl} do not predict
that.

\begin{table}[bhtp]
  \begin{center}
    \caption{Power law model $n=n_0(r/r_0)^{-\alpha}$ for the density structure of MonR2 IRS3.}
    \label{tab:monr2}
    \begin{tabular}{lc}

\hline \hline
\noalign{\smallskip}
Parameter & Value \\
\noalign{\smallskip}
\hline
\noalign{\smallskip}

Outer radius $r_0$ & 0.3 pc \\
Inner radius & 0.001 pc \\
$\alpha$     & 1.25 \\
$n_0$ (\hh)  & $1.21\times 10^4$ \ccm \\
CO/\hh\      & $8\times 10^{-5}$ \\
Dust mass    & 1.05 \msol \\

\noalign{\smallskip}
\hline

    \end{tabular}
  \end{center}
\end{table}

\subsection{Molecular abundances}

Table~\ref{tab:abs} lists the results of the radiative transfer
analysis, with the same source ordering as in Table~\ref{tab:cold}.
Abundances, if assumed constant, are $\sim$10$^{-8}$ for OCS,
$\sim$10$^{-9}$ for \hhs, \hhcs, SO and \soo, and $\sim$10$^{-10}$ for
HCS$^+$ and NS.  These trends are the same as found above for
beam-averaged column densities (\S~\ref{sec:cold}); the Monte Carlo
analysis does not change results by factors of more than a few.
Except for SO, \soo\ and OCS, the source-to-source spread in the
abundances is less than a factor of 10, and not related to
temperature.  These molecules (CS, \hhs, \hhcs, NS and HCS$^+$) appear
to trace the chemically inactive outer envelope ($T<100$~K), at least
in the observed transitions.  However, the high OCS abundances occur
all in warm sources.

\begin{table*}[hbtp]
  \begin{center}
    \caption{Molecular abundances and abundance ratios: outer envelope}
    \label{tab:abs}
    \begin{tabular}[lrrrrrrrrrrrrr]{lrrrrrrrrrrrrr}
\hline \hline
\noalign{\smallskip}
Source      & SO & SO$_2$ & OCS & CS$^a$ & H$_2$S & H$_2$CS & NS  & HCS$^+$ 
            
            & SO/ & \soo/ & SO/ & \hhs/ & NS/ \\
 &$10^{-9}$&$10^{-9}$&$10^{-9}$&$10^{-9}$&$10^{-9}$&$10^{-10}$&$10^{-11}$&$10^{-10}$
            & \soo & CS   & CS  & CS    & CS\\
\noalign{\smallskip}
\hline
\noalign{\smallskip}
NGC 7538 IRS9 & 1  & 0.5   &  2  & 10 &  3   & 10 & 1     & 5     & 2 &0.05&0.1&0.8& 0.001 \\
S140 IRS1     & 2  & 1     &$<$2 &  5 &  2   &  7 & 1     & 5     & 2 &0.2&0.4&0.8& 0.002 \\
W3 IRS5       & 10 & 10    & 0.5 &  5 &  3   &  3 & ...   & 0.2   & 1 & 2 & 2 & 2 & ... \\
MonR2 IRS3    & 5  & 1     & ... &  5 & $<$2 &  7 &$<$1   & ...   & 5 &0.2& 1 &$<$0.8& ... \\
AFGL 2136     & 1  & 0.5   & ... &  4 & $<$1 &  3 &$<$0.5 & 1     & 2 &0.1&0.25&$<$0.5&$<$0.001\\
W33A          & 1  & 1     & 20  &  5 &  3   & 13 & 2     & 2     & 1 &0.2&0.2& 2 & 0.004 \\
AFGL 2591     & 10 & 2     & 10  & 10 &  8   &  3 & 0.5   & 2     &10 &0.1& 1 & 2 & 0.0005 \\
NGC 7538 IRS1 & 5  & 1     &  2  & 10 &  3   & 10 & 1     & 2     & 5 &0.1&0.5&0.8& 0.001 \\
NGC 6334 IRS1 & 2  & 2     & 50  & 10 &  0.4 &  7 & 10    & 2     & 1 &0.2&0.2& 1 & 0.01 \\
\noalign{\smallskip}
\hline
\noalign{\smallskip}
Hot cores$^b$  & 4  & 20    & 5   &  8 & 9    & 6 & 80   & ...    &0.2 & 2 & 0.5 & 1 & 0.1 \\
PDRs$^c$       & 9  &0.1    & ... & 20 & 6    &...& ...  & ...    &100 &0.005&0.5&0.3& ...\\ 
Dark clouds$^d$& 20 & 4     & 2   &  1 & 0.8  & 6 & ...  & 0.6    & 5  & 4 & 20 & 0.8 & ... \\
Shocks$^e$     &200 &100    &10   &  4 & 4000 & 8 & ...  & 0.1    & 2  & 25& 50 &1000 & ... \\
Low mass protostars$^f$                                          
               & 4  &0.6    & 7   &  3 & 2    & 4 & ...  & 0.2    & 10 & 0.2& 1 & 0.7 & ... \\ 
\noalign{\smallskip}
\hline
    \end{tabular}

  \begin{tabular}{ll}

$^a$ From \citet{fvdt00}, except MonR2 IRS3 & $^d$ L134N: \citet{ohishi92} \\
$^b$ Source average from \citet{jh98sulf}   & $^e$ Orion Plateau: \cite{sutton95,minh90} \\
$^c$ Orion Bar: \citet{jansen95}            & $^f$ IRAS 16293-2422: \cite{fred02} \\

  \end{tabular}

  \end{center}
\end{table*}

\begin{figure*}[t]
\sidecaption
\includegraphics[width=14cm]{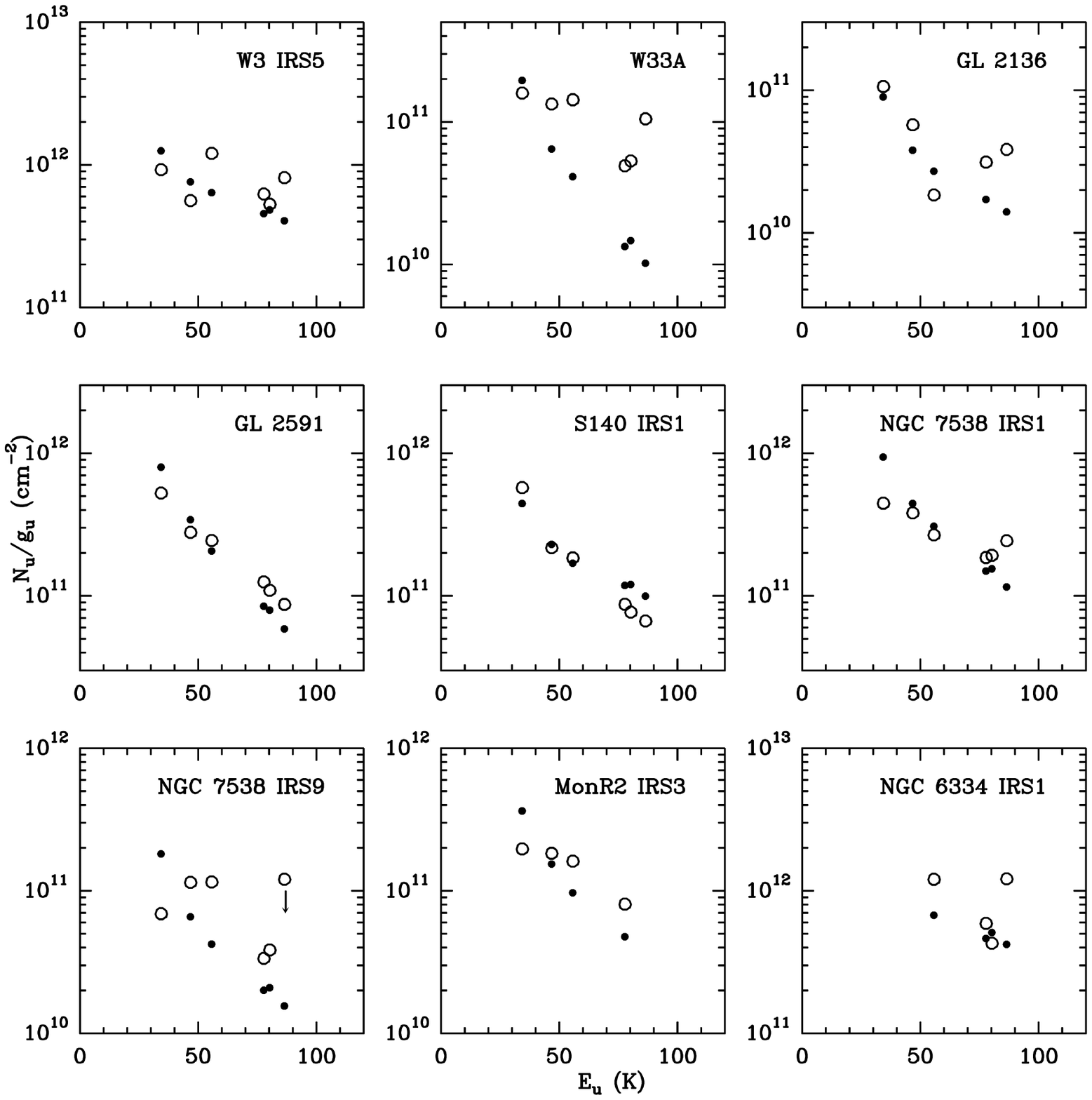}
    \caption{Observations of SO (open circles; envelope component only), and predictions
      from Monte Carlo models (filled dots). Line fluxes have been converted
      to upper state column densities following \cite{helm94}. The
      estimated error on the line fluxes is a factor of two, due to
      the uncertain contribution from high-velocity gas.}
    \label{fig:so_jump}
\end{figure*}

For SO and \soo, the number of detected lines is large enough to
investigate their abundances as a function of radius.
Figure~\ref{fig:so_jump} shows that the Monte Carlo models fit the SO
lines uniformly well over the observed range of energy levels for all
sources except W~33~A: there is no indication for changes in the SO
abundance between $T=30$ and $90$~K.  In the case of W~33~A, some of
the observed values appear to be higher than the models, which may be
due to an uncertain high-velocity contribution for these
lines.  Observations of higher-excitation SO lines ($N=10\to9$ near
430~GHz or $N=11\to10$ near 473~GHz) would be valuable to investigate
the effect of evaporating ice mantles on the abundance of SO.

\begin{figure*}[t]
\sidecaption
\includegraphics[width=14cm]{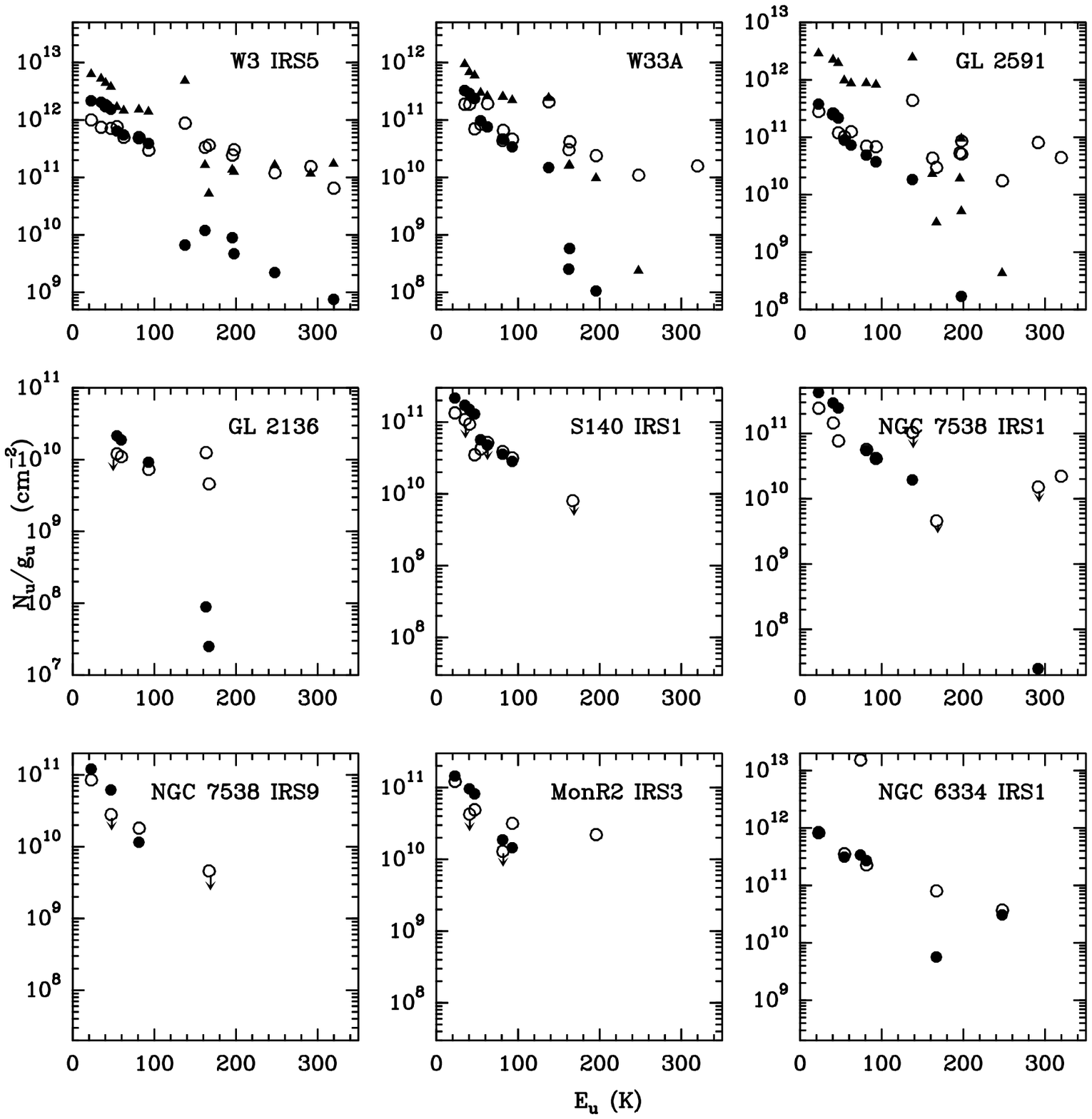}
    \caption{As previous figure, for \soo. Triangles indicate models
      with a temperature-dependent \soo\ abundance.}
    \label{fig:so2_jump}
\end{figure*}

Figure~\ref{fig:so2_jump} shows the results of the models for \soo.
Models with \soo/\hh\ $\sim 10^{-9}$ fit the lines with $E_u \ltsim
100$~K within factors of a few. However, these models underproduce the
lines with $E_u \gtsim 100$~K by factors of $\gtsim$100.  This
discrepancy is best seen for W3 IRS5, W33A and AFGL 2591, where the
most \soo\ lines are detected.  For these sources, we have run models
where the \soo\ abundance increases sharply (`jumps') when $T>100$~K.
These models are shown as triangles in Fig.~\ref{fig:so2_jump} and
quantified in Table~\ref{tab:jump}.  The \soo\ abundance increases by
a factor of 100 in the case of W3 IRS5 and W33A, and by a factor of
1000 for AFGL 2591.  These models reproduce all observed \soo\ lines
to factors $\sim$10.  The residuals are considerably bigger than the
expected error bars of factors of 2...3, mainly due to the uncertain
high-velocity contribution, which suggests that more detailed modeling
is needed. However, these `jump' models fit the data much better than
models where the \soo\ abundance is constant. The \soo\ abundances
derived from the high-excitation lines are similar to those based on
the infrared data, indicating that the same gas is probed. The data
for the other sources are consistent with similar abundance jumps, but
not enough high-excitation \soo\ lines have been detected to constrain
such models. However, the data of AFGL 2136, NGC 7538 IRS1 and MonR2
IRS3 allow estimates of \soo\ abundances in their inner envelopes,
which Table~\ref{tab:jump} also lists.
Except NGC~6334~IRS1, the jumps in the \soo\ abundance occur all in
sources with high $T$(\chhhoh), confirming that the abundance
enhancement is related to high temperatures, probably through
  ice evaporation.

\begin{table}[hbtp]
  \begin{center}
    \caption{Models with \soo\ abundance 'jumps'.}
    \label{tab:jump}
    \begin{tabular}{lrr}
\hline \hline
\noalign{\smallskip}

Source     & (\soo/\hh)$_{\rm out}$ & (\soo/\hh)$_{\rm in}$ \\
           & $10^{-9}$              & $10^{-7}$  \\
\noalign{\smallskip}
\hline
\noalign{\smallskip}
W3 IRS5      &   10                   & 10 \\
MonR2 IRS3   &    1                   & 12$^b$ \\
AFGL 2136    &   0.5                  &  0.8$^b$ \\
W33A         &    1                   &  1 \\
AFGL 2591    &    2                   &  2 \\
NGC 7538 IRS1&    1                   &  14$^b$ \\
\noalign{\smallskip}
\hline
\noalign{\smallskip}
IRAS 16293$^a$ &    0.5                 &  1 \\
\noalign{\smallskip}
\hline
    \end{tabular}
  \end{center}

$^a$ From \citet{fred02}.

$^b$ Estimated value.

\end{table}

The models discussed so far assumed a static envelope and aimed to
describe the low-velocity gas.  To investigate whether the
high-velocity gas could be due to infalling motions, we have modeled
the SO emission from W3 IRS5 under the assumption of gravitational
infall.  In this model, the infall velocity $V_{\rm inf}$ varies with
radius $R$ as $\sqrt{GM/R}$, where $G$ is the gravitational constant.
The luminosity of W3 IRS5, if due to a single star, suggests a stellar
mass of $\approx$30~\msol\ \citep{vacca96}, so that $V_{\rm inf}$
ranges from 0.45 to 10.66~\kms\ in the model. However, the synthetic
SO line profiles in 14--18$''$ beams have widths of 5--6~\kms\ FWHM,
much less than the observed value of 13.5~\kms. Since the other
sources have lower luminosities (hence infall speeds), it seems
unlikely that the high-velocity SO and \soo\ emission is solely due to
infall.

\section{Discussion}
\label{sec:disc}

\subsection{Comparison to other objects}
\label{sec:others}

Tables~\ref{tab:abs} and~\ref{tab:jump} also list molecular
abundances in other regions.  Only the case of the low-mass protostar
IRAS 16293 has been analyzed to the same level of detail as done here.
To reduce uncertainties introduced by methodological differences, the
table contains abundances scaled to the value of the commonly observed
CS molecule.  Using ratios also facilitates comparison with chemical
models, because these are independent of the total amount of sulphur
in the gas phase, which is not well-understood.  Ultraviolet
observations of diffuse clouds toward $\zeta$~Oph indicate that
sulphur is undepleted from its solar abundance of $\sim$2$\times
10^{-5}$ \citep{savage96}.  However, for dense clouds, it is unclear
what fraction of sulphur is locked up in dust grains.

The ratios observed in our sources differ considerably from those in
PDRs, dark clouds and shocks, indicating that in those regions, other
chemical processes dominate. The ratios in IRAS 16293 are at the low
end of the values found here, suggesting that in low-mass objects,
depletion of molecules onto grains is more important.

In the case of hot cores, both abundance ratios and absolute
abundances are similar to those observed here.  The chemical
similarity of our sources with hot cores prompts us to compare our
results with chemical models invoking ice evaporation.  Our
assumption is that source-to-source differences in abundances are due
to age differences, rather than to differences in initial physical or
chemical conditions.

\subsection{Comparison with chemical models}
\label{sec:chem_mdl}

The sulphur chemistry of hot cores, where ice mantles are evaporating
off dust grains and start a high-temperature chemistry, was studied by
\citet{char97}. The model assumes that the dominant form of sulphur in
the ices is \hhs, which after evaporation is destroyed by H and
\hhhop\ to form S and \hhhsp. Reactions with OH and \oo\ make first SO
and then \soo. Some \hhs\ reacts with C$^+$ into CS, which reacts with
OH to make OCS. The sulphur chemistry strongly depends on temperature
through its connection with the oxygen chemistry. At $T\gtsim 300$~K,
the \soo\ abundance remains factors of 10--100 below that of SO
because O and OH are driven into \hho. Results also depend on the
presence of \oo\ in the ice mantles.  Observational limits on solid
and gaseous \oo\ \citep{vdb99,gold00,pagani03} do not firmly rule out
either assumption, but do make the model without \oo\ ice more
plausible.  However, this model cannot directly be compared with our
observations, which probe a mixture of warm and cold gas.





The chemistry of protostellar envelopes was modeled by \citet{doty02}.
In the inner warm region, a hot core chemistry similar to that of
\citet{char97} is used, whereas in the cold outer parts, a
low-temperature chemistry including freeze-out is adopted.  The
initial conditions of the model depend on temperature to mimic the
effect of ice evaporation.  Although Doty et al.\ adopt the specific
temperature and density structure of AFGL~2591 as modeled by
\citet{fvdt00}, their results do not change by factors of more than a
few for our other sources, which have temperature and density
distributions similar to those of AFGL~2591.


The \citet{doty02} model has most gas-phase sulphur initially in S at
$T<100$~K (S/\hh=6$\times$10$^{-9}$) and in \hhs\ at $T>100$~K
(\hhs/\hh=1.6$\times$10$^{-6}$) to mimic the effect of ice
evaporation.  This choice of initial conditions is consistent with the
nondetection of the [S I] line at $25.249$~\mic\ toward our sources
with ISO.  The upper limit of $\tau<$0.01 in absorption implies $N$(S)
$<5\times 10^{15}$~\scm , so that for $N$(\hh) $\sim 10^{23}$~\scm,
S/\hh$<5\times 10^{-8}$, indicating that not all gas-phase sulphur in
the inner envelope can be in atomic form.  The inferred abundances
imply that SO$_2$ is one of the dominant sulphur-bearing gas-phase
molecules in the warm inner envelope.  However, the sum of all
sulphur-bearing molecules in the gas phase is still much lower than
the value derived for diffuse clouds (Savage \& Sembach 1996).  Unless
a major sulphur-bearing gas-phase species has been missed in this
survey, $\sim$90\% of the sulphur must be in a solid form which has
not yet been detected.

Infrared observations of our sources with ISO-SWS do not support the
assumption that \hhs\ is the main sulphur reservoir of the grain
mantles. Toward W33A, a source with large ice abundances, the
3.98~\mic\ band of solid \hhs\ is not detected to $\tau < 0.02$, which
using a band strength of $5\times 10^{-18}$ cm~mol$^{-1}$ gives
$N_s$(\hhs) $<3.6\times 10^{17}$~\scm, or $N_s$(\hhs)/$N_s$(\hho)
$<0.03$ (W.~Schutte, priv.\ comm.). However, solid OCS was detected
with an abundance of $N_s$(OCS)/$N_s$(\hho) $=(0.4-1)\times 10^{-3}$
or OCS/\hh\ $=(4-14)\times 10^{-8}$ \citep{palum97}.  Evaporation of
solid OCS may explain our measured gas-phase OCS abundances which are
otherwise unaccounted for.  W~33~A is the only source in our sample
for which both gas-phase and solid OCS have been detected.  The
results imply a gas/solid ratio of $\sim$0.5, much higher than that
found for e.g. H$_2$O and CO$_2$ in this source
\citep{mieke:co2,mieke:h2o}.  Observations of both gas-phase and
solid-state OCS toward other sources are needed to determine the role
of grain-mantle evaporation.  Observations of OCS lines from a wide
range of energy levels may reveal an abundance increase at $T\gtsim
100$~K, as indeed found for IRAS 16293 \citep{fred02}. The low \txc\
of \hhs\ and the high value of OCS (\S~\ref{sec:texc}) support these
conclusions.

The effect of evaporating OCS-rich ice mantles on hot core chemistry
is explored by \citet{doty03}. Briefly, the gas-phase abundances of
CS, HCS$^+$ and OCS increase about linearly with the fraction of solid
OCS, while those of \hhs, SO and \soo\ decrease about linearly. The
\hhcs\ molecule is not affected much. Since the gas-phase chemistry of
CS, HCS$^+$ and OCS depends only weakly on time, the use of sulphur
molecules as chemical clock diminishes when \hhs\ is only a minor
sulphur carrier in the ice mantles. However, the 'jump' models for
\soo\ (Table~\ref{tab:jump}) indicate that either \hhs\ or \soo\ is
present in the grain mantles. The limits on solid \soo\ from infrared
observations are not very stringent \citep{boog98}.


We have compared our observations to the models of \citet{doty02},
modified to have OCS in the evaporating ice mantles (S.\ Doty, priv.\ 
comm.).  This model reproduces our observed abundances of CS, SO and
\hhcs\ in the outer envelopes (Table~\ref{tab:abs}) to within factors
of a few for chemical ages of $\sim$10$^5$~yr. The abundances of \soo\ 
in the inner envelopes (Table~\ref{tab:jump}) are also reproduced for
the same chemical age. The model abundances of \hcsp\ and \soo\ in the
outer envelopes are factors of $\sim$10 below the observed values,
which may be due to the adopted initial conditions, or the chemical
network, which is based on the UMIST reaction rate database
\citep{umist99}\footnote{http://www.rate99.co.uk}.  The outer envelope
abundances of \hhs\ and OCS are underproduced even more (factors of
$\sim$100).  However, the small number of observed lines forced us to
assume constant abundances for these species, while they may in fact
be confined to the inner envelopes.



\subsection{The role of shock chemistry}
\label{sec:shock_mdl}

As an alternative to ice evaporation models, we consider the effect of
non-dissociative shocks on the molecular envelopes surrounding the
protostars. Such shocks occur by the interaction of the molecular
outflows of protostars with their envelopes.  Shocks fast enough to
dissociate \hh\ ($v\gtsim 40$~\kms) would have a major impact on their
molecular composition which our data do not indicate.  Although our
sources show outflows faster than that, both in infrared absorption
\citep{mitch91:episode} and in submillimeter emission
\citep{mitch91:irs9,mitch92} of CO, their filling factor must be
small.



Our observed SO, \soo\ and \hhs\ abundances in the outer envelopes are
consistent with those in postshock gas if sulphur was initially in
molecular form \citep{leen88}. The observed similarity of these
abundances is not reproduced by the \citet{gpdf93} model which has
most sulphur initially in atomic form. However, the two models are
hard to compare because one is a two-fluid treatment, while the other
treats neutral and charged particles as one fluid. Before concluding
that postshock gas reproduces our data, we recommend that a two-fluid
calculation is performed with most sulphur initially in molecular
form.

Two other molecules suggest that the envelopes of young high-mass
stars may have been processed by shocks, although neither one
conclusively. First, ISO-SWS data show low ratios of gas-phase to
solid-state \coo\ ($\ltsim 0.1$; \citealt{evd98}) and the \coo\ gas
phase abundance remains low through the 100--300~K temperature regime.
After evaporating off grains, \coo\ must be promptly destroyed, which
shocks can do in reactions with H \citep{char00} or perhaps \hh\ 
\citep{doty02,talbi02}. Second, \citet{hatch02} measured NS/CS ratios of
0.02-0.05 in a sample of six hot cores and interpreted these as
evidence for shocks. The main reactions to form NS require SH and NH
which are produced from OH.  Shocks use OH to form \hho\ and suppress
the production of NS.  The values of NS/CS=0.001 -- 0.01 measured here
may indicate that shocks play a role too. However, for $t=3\times
10^4$~yr, the \citet{doty02} model also predicts NS/CS $\sim 10^{-3}$,
so this ratio cannot be used to demonstrate the influence of shocks.

\section{Conclusions}
\label{sec:concl}

Submillimeter observations of SO, \soo, \hhs, \hhcs, OCS, NS and
HCS$^+$ toward nine embedded massive stars show that:

\begin{enumerate}
\item High-velocity gas contributes 10--50\% to the CS, SO and \soo\ 
  emission in $15-20''$ beams. This gas may either trace the outflows
  which are also seen in CO, or they may trace infall motions.
\item The \soo\ abundance increases from dark cloud levels in the
  outer envelope ($T<100$~K) to levels seen in hot cores and shocks in
  the inner envelope ($T>100$~K).
\item Molecular abundances are consistent with a model of ice
  evaporation in an envelope with gradients in temperature and density
  for a chemical age of $\sim$10$^5$~yr. 
\item The high observed abundance of OCS, the fact that \txc
  (OCS)$\gg$\txc (\hhs), and the detection of solid OCS and limit on
  solid \hhs\ all suggest that OCS is a major sulphur carrier in grain
  mantles, rather than \hhs.
\item For most other sulphur-bearing molecules, the source-to-source
  abundance variations by factors of up to 10 do not correlate with
  previously established evolutionary trends in temperature tracers.
  These species probe the chemically inactive outer envelope.  Our
  data set does not constrain the abundances of \hhs\ and SO in the
  inner envelope, which, together with \soo, are required to use
  sulphur as a clock.
\end{enumerate}

These conclusions could be followed up by, respectively:
\begin{enumerate}
\item interferometric observations at $\ltsim 1''$ resolution of SO
  lines from a wide range of energy levels.
\item similar interferometric observations of \soo.
\item updating the model of \citet{leen88} using a two-fluid
  treatment, and extending it to later times.
\item a multi-line study of OCS, covering energy levels well above and
  well below 100~K.
\item observations of high-excitation SO and \hhs\ lines, e.g.\ SO
  430~and 473~GHz and \hhs\ 369~GHz.
\end{enumerate}

Items 3 and 4 are within the capabilities of current computational and
observational facilities. The CSO and JCMT may be able to carry out
item 5 in very good weather, but the future Atacama Pathfinder
Experiment (APEX) can benefit from an even better site. Current
interferometers are not sensitive enough for items 1 and 2, which
require observing several lines and several sources. However, future
instruments such as the Combined Array for Research into Millimeter
Astronomy (CARMA) and the Atacama Large Millimeter Array (ALMA) will
be well-suited for these projects.

\begin{acknowledgement}
  The authors thank the JCMT and IRAM staffs for their support,
  Malcolm Walmsley, Holger M\"uller, Jennifer Hatchell and Xander
  Tielens for useful discussions, and Willem Schutte, Fred Lahuis and
  Steve Doty for input. Astrochemistry in Leiden is supported by the
  Nederlandse Organisatie voor Wetenschappelijk Onderzoek (NWO)
  through grant 614-41-003 and a Spinoza award.
\end{acknowledgement}

\bibliographystyle{aa}
\bibliography{so2}
 
\end{document}